\documentclass[11pt]{article}
\usepackage{amssymb,amsmath,cite,geometry,graphicx,url,mathrsfs}
\catcode`\@=11

\@addtoreset{equation}{section}
\def\theequation{\arabic{section}.\arabic{equation}}
     
\catcode`\@=11
\def\thesection{\arabic{section}}

\def\appendix{\setcounter{section}{0}
        \def\thesection{Appendix.}
        \def\theequation{\Alph{section}.\arabic{equation}}}
\def\section{\@startsection{section}{1}{\z@}{3.5ex plus 1ex minus
   .2ex}{2.3ex plus .2ex}{\large\bf}}
     
\long\def\@makefntext#1{\parindent 0cm\noindent
\hbox to 1em{\hss$^{\@thefnmark}$}#1}

\newcommand{\captionfonts}{\small}
\makeatletter  
\long\def\@makecaption#1#2{%
  \vskip\abovecaptionskip
  \sbox\@tempboxa{{\captionfonts #1: #2}}%
  \ifdim \wd\@tempboxa >\hsize
    {\captionfonts #1: #2\par}
  \else
    \hbox to\hsize{\hfil\box\@tempboxa\hfil}%
  \fi
  \vskip\belowcaptionskip}
\makeatother   

\begin{document}
\begin{titlepage}
\vspace{.5in}
\begin{flushright}
November 2021\\
 \end{flushright}
\vspace{.5in}
\begin{center}
{\Large\bf
 Spacetime foam, midisuperspace,\\[1ex]
  and the cosmological constant}\\  
\vspace{.4in}
{S.~C{\sc arlip}\footnote{\it email: carlip@physics.ucdavis.edu}\\
       {\small\it Department of Physics}\\
       {\small\it University of California}\\
       {\small\it Davis, CA 95616}\\{\small\it USA}}
\end{center}

\vspace{.5in}
\begin{center}
{\large\bf Abstract}
\end{center}
\begin{center}
\begin{minipage}{4.65in}
{\small Perhaps the cosmological constant really is huge at the Planck scale, but is ``hidden'' by Planck 
scale quantum fluctuations of spacetime.  I briefly review this proposal and provide some evidence, 
coming from a simplified midisuperspace model, that an appropriate ``foamy'' structure can do the job 
of hiding a large cosmological constant, and can persist under time evolution. 
 }
 \end{minipage}
\end{center}
\end{titlepage}
\addtocounter{footnote}{-1}
\section{Spacetime foam and the cosmological constant \label{foam}}

Gravity is universal: it couples with equal strength to all forms of energy.  This principle,
a version of the principle of equivalence, lies at the foundation of general relativity, and in ordinary
settings it is exquisitely well tested \cite{tests}.  But ``all forms of energy'' should presumably include 
vacuum energy, the energy of quantum fluctuations in empty space, whose gravitational 
interactions should take the form of a cosmological constant $\Lambda$.  We do not really know how 
to calculate this energy, but standard effective field theory methods yield a value some sixty 
orders of magnitude higher than observational limits \cite{Martin,EFT,Carroll}.
 
 Understanding this ``cosmological constant problem'' is a notoriously difficult task \cite{Weinberg}.  But
 vacuum energy comes from vacuum fluctuations, and it is possible that a solution to the  
problem might come from the same place.  More than sixty years ago, Wheeler suggested that
\cite{Wheelerb}
\begin{quote}
``\dots it is essential to allow for fluctuations in the metric and gravitational interactions in any
proper treatment of the compensation problem---the problem of compensation of `infinite' energies that
is so central to the physics of fields and particles.''
\end{quote}
What if this is right?

A cosmological constant is the darkest of dark energy, detectable only through its gravitational effects.
Gravity, in turn, is the curvature of spacetime.  So one way to interpret Wheeler's proposal is to ask
whether very high curvature at the Planck scale can somehow be hidden from view at larger
distances.

If spacetime were Riemannian---that is, if the metric were positive definite---this would be easy.  At
large scales, a golf ball has a curvature of about of about .2\,cm${}^{-2}$.  But at small scales, the
dimples have a curvature some 500 times greater, with regions of positive and negative curvature 
averaging out at larger distances.  This is commonplace; in Wheeler's analogy, the
surface of the ocean looks flat from an airplane, but up close one can see complex 
geometry, with ``foam forming and breaking, breaking and forming''  \cite{Wheeler3}.
For a spacetime, with its Lorentzian metric, things are a bit more complicated.  Here, a cosmological 
constant normally manifests itself as an accelerating expansion or contraction.  But in this setting, 
too, perhaps cancellations between expanding and contracting regions can occur.

A concrete realization of this idea was suggested in \cite{Carlip1}.  Let us start with  a 
``typical'' spatial slice $\Sigma$ at constant time.  The geometry of such a slice is  characterized by 
an intrinsic metric $q_{ij}$ and an extrinsic curvature $K^i{}_j$ (or equivalently a metric and its conjugate 
momentum $\pi^i{}_j$).  Physically, the mean curvature $K=K^i{}_i$ is the local expansion, that is, the
local Hubble constant, while the traceless part $\sigma^i{}_j$ of $K^i{}_j$ is the local shear;
when we say we are observing the cosmological constant, this is usually what we are really looking at.

This geometric data cannot be chosen arbitrarily, though.  It must obey a set of constraints, the 
Hamiltonian constraint
 \begin{align}
&{}^{\scriptscriptstyle(3)}\!R + K^2 - K^i{}_jK^j{}_i - 2\Lambda = 0 \label{a1a}\\ 
\intertext{and the momentum constraint}
&D_i(K^i{}_j - \delta^i_jK) = 0 ,
\label{a1b}
\end{align}
where $D_i$ is the covariant derivative compatible with $q_{ij}$.  Note that this data has a time 
reversal symmetry: if $(q_{ij} ,K^k{}_\ell)$ satisfies the constraints, so does $(q_{ij},-K^k{}_\ell)$

Suppose, as Wheeler proposed,  that $\Sigma$ is ``foamy,'' having a complicated geometric 
and topological structure near the Planck scale.  A classical result of three-manifold topology tells us 
that there is a nearly unique\footnote{For orientable three-manifolds, the prime decomposition is unique; 
for nonorientable manifolds, there is one free choice.} decomposition of the slice into elementary 
topological pieces, called prime manifolds \cite{Milnor,Giulini}.  These elements are attached by 
connected sums; to a physicist, this simply means they are joined by wormholes.  More recently,
Chrusciel, Isenberg, and Pollack \cite{Chrusciel,Chruscielb} showed that this ``gluing'' process can be
made to respect the constraints: nearly any pair of three-manifolds with metrics and extrinsic curvatures 
satisfying (\ref{a1a})--(\ref{a1b}) can be joined by a wormhole in a way that continues to satisfy the
constraints, while changing the geometric data only in arbitrarily small regions around the wormhole 
mouths.

Now suppose that  spacetime foam exists, and that quantum gravity has no preferred direction of 
time.  Start with random collection of three-manifolds $\Sigma_1,\Sigma_2,\dots,\Sigma_N$ with initial 
data $(g_\alpha,K_\alpha)$.  Join these to form a large, perhaps topologically complex, ``foamy'' manifold
\begin{align} 
{\widetilde\Sigma} = \Sigma_1\#\Sigma_2\#\dots\#\Sigma_N  ,
\label{a2}
\end{align}
where $\#$ denotes a connected sum and where the geometric data is glued in the manner of 
\cite{Chrusciel,Chruscielb}.  Since there is no preferred direction of time, for each factor $\Sigma_\alpha$
the data $(g,K)$ and $(g,-K)$ are equally likely.  Then for any reasonable definition of averaging,
\begin{align}
\langle K^i{}_j\rangle \sim 0
\label{a3}
\end{align}
for a large enough collection of elementary factors.  Note that one cubic centimeter has about 
$10^{100}$ Planck volumes, so for Planck-scale foam the number of factors in such an average is enormous.
We thus have a large class of initial data describing spacetimes whose average expansion and shear 
vanish, even if the cosmological constant is much greater than zero.

How typical is this situation? We don't yet know.  For a fixed spatial topology, the specific connected sum 
geometries of  \cite{Chrusciel,Chruscielb} are probably fairly special, but there are also other kinds of initial 
data in which positive and negative expansions cancel.   For topological fluctuations, the connected sum 
decomposition always exists, and time-reversed geometries exist on each prime factor.   More physically, 
local fluctuations should at least arguably have an arbitrary sign of $K$---there is no obvious reason to 
impose a global direction of time on a local quantum fluctuation---so if the usual spacetime foam picture of 
quasi-independent Planck scale quantum fluctuations is correct, one might reasonably expect a similar 
cancellation.

\section{Evolution}

So far, I have focused on the geometry of a single spatial slice.  A crucial question  is
whether an initial foamy structure is preserved by evolution.  There are two natural guesses.
On the one hand, expanding regions should become larger over time, eventually dominating.  On
the other hand, there was nothing special about the choice of the slice $\Sigma$, so if ``foam''
is typical, perhaps the foamy structure should reproduce itself, with expanding regions themselves filling 
up with expanding and contracting bubbles.

A number of attempts have been made to answer this question classically \cite{Carlip1,Wang,%
Carlipreply,Tsamis,Buchert}, with inconclusive results.\footnote{A related line of research asks about the
asymptotic behavior of spacetimes with a positive cosmological constant \cite{Kleban,Mirbabayi,Moncrief,%
Creminelli}.  Partial results exist, but the work so far assumes a slice on which the expansion is constant, or at least
everywhere positive, and thus does not apply to the foamy spacetimes discussed here.}   What has become
clear, though, is that the question ultimately requires a quantum mechanical answer.  Classically, initial data 
with local nontrivial topology always evolves to form singularities \cite{Gannon,Lee}, as does the connected 
sum data of Chrusciel et al.\ \cite{Burkhartb}.  If we hope to have a sensible global description, we cannot
avoid quantum effects.

This would be a hard task even if we had a complete quantum theory of gravity with which to attempt 
the calculation.  But while such a theory does not yet exist, it is widely (although perhaps not universally) 
believed that any final quantum theory of gravity will include some form of the Wheeler-DeWitt equation 
\cite{DeWitt}, at least as an approximation.  This equation is the quantum version of the constraints,
obtained by rewriting (\ref{a1a})--(\ref{a1b}) as operator expressions acting on the wave function, 
with the usual canonical quantization rule
\begin{align}
&\pi^{ij} \rightarrow \frac{\hbar}{i}\frac{\delta\ }{\delta q_{ij}}   ,
\label{b1a}
\intertext{where the canonical momentum is}
&\pi^{ij}  =  -\frac{1}{2\kappa^2}\sqrt{q}(K^{ij} - q^{ij}K) . \label{b1b}
\end{align}
(I use the conventions of \cite{Carlipbook}.)  The full Wheeler-DeWitt equation is still far too complicated
to be tractable, but I will argue below that we can learn something from simplified 
midisuperspace models.  

First, though, we must confront  a general issue, the notorious ``problem of time'' in quantum 
gravity \cite{Kuchar}.
Ordinary quantum field theory takes place in a fixed spacetime, and while there are some subtleties
in the choice of time slicing \cite{Varadarajan}, this background provides a setting in which
to define time evolution.  In quantum gravity, on the other hand, spacetime is dynamical, and there is no 
preferred choice of time coordinate.  One must instead construct ``time'' from physical observables, 
treating evolution as relational.   There are a number of attempts to do this, but the standard choices---volume
as time \cite{vol}, extrinsic curvature as time \cite{York}, mean curvature flow \cite{Kleban}, cosmological
time \cite{cos}---require spacetimes that are expanding everywhere, and simply don't apply to the geometries
considered here.  An alternative, proposed by Brown and Kucha{\v r} \cite{Brown,Husain}, is to introduce 
a cloud of ``clocks,'' noninteracting particles whose proper time $T$ can be used to determine 
evolution.  This is not ideal---the clocks back-react on the spacetime, and one must restrict to quantum 
states in which this effect is small---but I do not know an alternative.

With this choice of time, the Wheeler-DeWitt equation becomes
\begin{align}
&i\sqrt{q}\,\frac{d\Psi[q]}{dT} =  \left(\frac{\hbar^2\kappa^2}{\sqrt{q}}
    G_{ijkl}\frac{\delta\ }{\delta q_{ij}}\frac{\delta\ }{\delta q_{kl}}
    - \frac{1}{2\kappa^2}\sqrt{q}\,({}^{\scriptscriptstyle(3)}\!R -2\Lambda)\right)\Psi[q]  , \label{b2a}\\
&D_i\frac{\delta\Psi[q]}{\delta q_{ij}} = 0 , \label{b2b}
\end{align}
where $G_{ijkl} = q_{ik}q_{jl} + q_{il}q_{jk} - q_{ij}q_{kl}$ is the DeWitt metric 
on the space of metrics. 

\section{Midisuperspace}

If we could solve the Wheeler-DeWitt equation---and find the correct inner product and a suitable
set of observables---we would have a quantum theory of gravity.  We cannot.  But there are  
simplified settings in which much more progress can be made: minisuperspaces, in which all but
a finite number of geometric degrees of freedom are frozen out, and midisuperspaces, in which
most of the geometric degrees of freedom are frozen out but an infinite number are retained.
As I showed in \cite{Carlipmidi}, a particular midisuperspace, the space of locally spherically 
symmetric metrics, provides a good model for the foamy spacetimes of section \ref{foam},
including a connected sum construction that contains both expanding and contracting regions.

The classical properties of locally spherically symmetric spacetimes were investigated extensively 
about 25 years ago \cite{Witt1,Witt2,Witt3,SW,Bengtsson}.  Birkhoff's theorem tells us that every
point in such a spacetime has a neighborhood isometric to some region of Schwarzschild-de
Sitter space, where for concreteness I will assume $\Lambda>0$.  But as Morrow-Jones and
Witt showed \cite{Witt1}, patches of Schwarzschild-de Sitter can be glued together to build
spacetimes with far more complicated geometries and topologies.

Classically, the spatial metric of a  locally spherically symmetric spacetime on a constant time slice
takes the form
\begin{align}
ds^2 = h^2d\psi^2 + f^2(d\theta^2 + \sin^2\theta\, d\varphi^2) ,
\label{c1}
\end{align}
where $h$ and $f$ are functions of $\psi$ and $t$.  As shown in \cite{Witt1}, the classical constraints
can be solved exactly to determine the two independent components of the extrinsic curvature,
$K^\psi{}_\psi$ and $K^\theta{}_\theta$, in terms of $f$ and $h$.  The solution depends on a single
integration constant $\gamma$, essentially a black hole mass, and is then unique up to sign.  
Patches of different geometries $(h,f,\pm K^\psi{}_\psi, \pm K^\theta{}_\theta)$ can then be glued
together by connected summation, in a process closely analogous to that of section \ref{foam}, 
leading again to a manifold with the structure (\ref{a2}).

In \cite{Carlipmidi}, I considered the simplest case, a manifold with the spatial topology $S^1\times S^2$.
For the metric, this simply means imposing periodicity in $\psi$.  Topologically, such a manifold 
is formed by starting with a solid three-ball, cutting out a ball at the center to form a space $[0,1]\times S^2$,
and then identifying the inner and outer boundaries $\{0\}\times S^2$ and $\{1\}\times S^2$.  The advantage
of this construction is that it is easily generalized: instead of starting with a single thick shell $[0,1]\times S^2$,
we can take a sequence of concentric thin shells 
$$[0,\psi_1]\times S^2,\,[\psi_1,\psi_2]\times S^2,\,\dots,[\,\psi_N,1]\times S^2$$
and join them in an onion-like structure by connected summation.  I showed in \cite{Carlipmidi} that
as long as the integration constant $\gamma$ obeys a suitable inequality, this construction allows a
mixture of shells of positive and negative expansion.

This construction thus mimics that of Chrusciel, Isenberg, and Pollack \cite{Chrusciel,Chruscielb},
but in a setting simple enough that one can treat the Wheeler--DeWitt equation seriously.  Eqn.\ 
(\ref{b2a}) becomes a Schr{\"o}dinger-like equation
\begin{align}
i\frac{d\Psi}{dT} = \left[ \frac{3\kappa^2}{64\pi^2}\left(\frac{1}{f^2}\frac{\delta\ }{\delta h}\right)^2 
    + \frac{3}{\kappa^2}\frac{1}{f^2}\left(\frac{f^{\prime 2}}{h^2} -1\right) + \frac{\Lambda}{\kappa^2}
    + \frac{3\gamma}{\kappa^2 f^3}\right]\Psi ,
\label{c2}
\end{align}
while the momentum constraint (\ref{b2b}) can essentially be solved, and tells us that wave functions 
should be built from integrals of the form
\begin{align}
F[h,f] = \int\!d\psi\,h L[f,Df,D^2f,\dots] \qquad\hbox{with $\displaystyle D = \frac{1}{h}\frac{d\,}{d\psi}$} .
\label{c3}
\end{align}

\section{WKB}

We can now look for stationary states in the WKB approximation,
\begin{align}
\Psi[f,h;T] = {\tilde\Psi}[f,h]e^{-iET}  \quad\hbox{with ${\tilde\Psi} = A e^{iS}$} .
\label{d1}
\end{align}
The lowest order equation for $S$ can be solved exactly, giving 
\begin{align}
S = \frac{8\pi}{\kappa^2}\int\!d\psi\,\sigma[h,f;\psi]ff'\left\{\sqrt{1+\beta h^2} - \tanh^{-1}\sqrt{1+\beta h^2} \right\} ,
\label{d2}
\end{align}
where
\begin{align}
\beta = \frac{f^2}{f^{\prime2}}\left(\frac{\tilde\Lambda}{3} - \frac{1}{f^2} + \frac{\gamma}{f^3}\right)
\quad \hbox{with ${\tilde\Lambda} = \Lambda - \kappa^2E$}
\label{d3}
\end{align}
and where $\sigma$ is a functional of $h$ and $f$ and a function of $\psi$ such that
\begin{align}
\left\{ \begin{array}{ll}\sigma^2 = 1 \qquad&\hbox{almost everywhere}\\[.5ex]
  \partial_\psi\sigma =0 &\hbox{unless}\  1+\beta h^2 = 0  .
   \end{array}\right.
\label{d4}
\end{align}
Geometrically, $\sigma$ is the sign of the expansion.  As in the classical case, it is not determined
by the constraints, and can change between layers, jumping when $1+\beta h^2$ passes through zero.

The Wheeler-DeWitt equation (\ref{c2}) involves two functional derivatives at a single point, leading to a divergence,
and the equation for the amplitude $A$ requires regularization.  But with a standard heat kernel regularization
\cite{Horiguchi}, we again have a closed form solution,
\begin{align}
A=\exp\left\{ \frac{\alpha}{2}\int\!d\psi\,\frac{1}{\sqrt{\beta}}\tan^{-1}(\sqrt{\beta}h)\right\}  ,
\label{d5}
\end{align}
where $\alpha$ is a regularization parameter with dimensions of inverse length, which is plausibly
the inverse length of the $S^1$ factor in the spatial geometry.

So we have solutions of the Wheeler-DeWitt equation; now we must interpret them.  Our wave
functions depend on the spatial metric $(f,h)$, and a physical interpretation will have to describe
how probabilities vary over this space of three-metrics.  As shown in \cite{Carlipmidi}, 
the solutions (\ref{d2})--(\ref{d5}) have several key features:
\begin{enumerate}
\item There are de Sitter-like regions, in which $f$ is large and $\sigma$ is fixed at either $+1$
or $-1$.  In these regions, the expansion is large, with a sign determined by $\sigma$.  But as long as
\begin{align}
0 < \gamma < \frac{2}{3\sqrt{\tilde\Lambda}} ,
\label{d6}
\end{align}
the wave function also has support on multilayered ``onion-like'' regions, where the sign of the
expansion changes each time $\sigma$ changes sign.
\item The de Sitter-like regions, and more generally regions with large expansion, have probabilities
that are strongly, although not exponentially, suppressed.  Probabilities are enhanced both for regions
in which the sign of the expansion changes and for ``nearby'' regions where the expansion is small.
\item For most of the configuration space, the solutions appear to be genuinely nearly time independent:
the small average expansion coming from ``foamy'' cancellations is preserved under evolution.
\end{enumerate}

This last point is crucial, and deserves further explanation.  The states discussed here are stationary,
so it is trivially true that probabilities are time independent.  But as we know from the ordinary WKB
approximation in quantum mechanics, this can be misleading.  The WKB wave function for a particle
reflecting off a potential barrier, for instance, is stationary, but the physical process clearly is not.  Fortunately, 
there is a simple diagnostic for such a situation: the probability \emph{current} can reveal the flow of
probability, and through that the hidden dynamics.

For quantum gravity, even in a Schr{\"o}dinger-like picture like the one used here, it is difficult to write 
down a diffeomorphism-invariant probability current.  But in \cite{Carlipmidi}, modes are constructed
that are at least invariant under spatial diffeomorphisms.  They reveal the hoped-for behavior: 
de Sitter-like regions show a clear probability flow, but in multilayered foamy regions the current
becomes extremely small.

There are, of course, still many open questions, even at the level of the simple midisuperspace model.
Perhaps the most important involves the inner product.  We know
that the inner product on the space of solutions of the Wheeler-DeWitt equation must be
gauge fixed, and this process can influence the probability measure \cite{Woodard}.  Some progress 
toward understanding this measure was made in \cite{Carlipmidi}, where is was shown that it
preserves the symmetry between expanding and contracting solutions and thus
does not prevent the cancellations in foamy spacetimes.  But further work is certainly
needed.

As usual in quantum gravity, we also have a shortage of good observables.  The extrinsic
curvature, for instance, is a natural indicator of expansion, but it is not diffeomorphism
invariant, and is not a good quantum observable.  Quite generally, observables in quantum
gravity must be nonlocal \cite{Torre}, and such objects are hard to construct and interpret.

It should also be possible to look at more general topologies.  The three-sphere $S^3$ should 
be fairly easy, but more complicated topologies might also be tractable.  What would really be interesting 
would be to look at the possibility of topology change, restoring Wheeler's idea of spacetime foam as
fluctuating topology as well as fluctuating geometry.  In particular, one might ask whether 
contracting ``bubbles'' will nucleate in an expanding region, and expanding ``bubbles'' in a
contracting region.  This would call for a study of Euclidean instantons in our midisuperspace, 
perhaps a feasible project, but one for the future.

\begin{flushleft}
\large\bf Acknowledgments
\end{flushleft}

I would like to thank Andy Albrecht, Sean Carroll, Piotr Chrusciel, Markus Luty, Don Page, 
Daniel Pollack, Albert Schwarz, and Bob Wald for helpful discussions. I would also like to thank 
the Quantum Gravity Unit of the Okinawa Institute of Science and  Technology (OIST), where part 
of the work was completed, for their hospitality.   This work was supported in part by Department
of Energy grant DE-FG02-91ER40674.

\end{document}